\documentclass[aps,prb,twocolumn,preprintnumbers,amsmath,amssymb,floatfix]{revtex4}

\usepackage{graphicx}
\usepackage{longtable}
\usepackage{bm}
\usepackage{multirow}
\usepackage{hyperref}
\usepackage{amssymb}
\usepackage{amsmath}
\usepackage{array}

\begin{document}

\title{Isospin-phonon coupling and Fano-interference in spin-orbit Mott insulator Sr$_{2}$IrO$_{4}$ }

\author{K. Samanta}
\author{D. Rigitano}
\author{P. G. Pagliuso}
\author{E. Granado}

\affiliation{\textit{``Gleb Wataghin'' Institute of Physics, University of Campinas - UNICAMP, Campinas, S\~ao Paulo 13083-859, Brazil}}

\begin{abstract}

The \textit{isospin-phonon} coupling in Sr$_{2}$IrO$_{4}$ is investigated by temperature dependent phonon Raman scattering. Anomalous behavior in the frequency of all studied optical phonons is observed below the magnetic transition temperature T$_{N} \sim$ 240 K. The strongest effect is detected for the A$_{1g}$ mode at 272 cm$^{-1}$ associated with the modulation of the Ir-O-Ir bond angle. Additionally, the A$_{1g}$ mode at 560 cm$^{-1}$ shows a Fano asymmetric lineshape sensitive to T$_{N}$, supporting the existence of low-energy ($\sim$70 meV) electronic excitations that are renormalized by the magnetic order. These results reveal new aspects of the interaction between the crystal lattice and electronic degrees of freedom in this \textit{spin-orbit} entangled Mott insulator.     

\end{abstract}

\maketitle

Materials with unique combination of \textit{spin-orbit} coupling (SOC) and electron correlations offer opportunities for the discovery of novel quantum states of matter\cite{Pesin2010}. The Ruddlesden-Popper series of layered perovskite strontium iridates (Sr$_{n+1}$Ir$_{n}$O$_{3n+1}$, n = 1, 2) are the key materials in this class. Mutual cooperation of strong SOC ($\sim$ 0.5 eV) and moderate electron correlation (U $\sim$ 1.5 - 2 eV) energies give rise to the emergence of the \textit{spin-orbit} entangled insulating state in Sr$_{2}$IrO$_{4}$. The SOC creates a half filled band of isospins (J$_{eff}$ = 1/2), which is further split by the on-site Coulomb interaction and opens up a Mott-like insulating gap\cite{Kim2008,Kim1329,Jackeli2009,Pesin2010,Arita2012,Kim2012}. The parent Sr$_{2}$IrO$_{4}$ crystallizes in a tetragonal phase with space group \textit{I4$_{1}$/acd}, showing a rotation of the IrO$_{6}$ octahedra by 11$^{\circ}$ along the \textit{c}-axis\cite{Crawford1994,HUANG1994355}. The presence of a significant Dzyaloshinskii-Moriya-type (DM) exchange interaction leads to a canted magnetic structure for the isospins with a weak ferromagnetic moment of 0.06-0.14 $\mu_{B}$/Ir below T$_{N}$ $\sim$ 240 K\cite{Kim1329,Cao1998,Chikara2009,Liu2015}.

The effect of cross-coupling between collective spin and orbital ordering with the lattice degrees of freedom has an important role on the fundamental properties of correlated materials. Particularly, the \textit{spin-lattice} coupling strongly depends on \textit{spin-orbit} interaction, and it is important to stabilize the uncommon magnetic behavior of Sr$_{2}$IrO$_{4}$\cite{Porras2019}. The external influences like temperature, application of pressure, and chemical doping can change the bond length and/or bond angle, which leads to a change in magnetic ordering and consequently to the \textit{spin-lattice} coupling\cite{Lee2011}. In Sr$_{2}$IrO$_{4}$, the strong SOC plays a crucial role in the realization of the magnetic behavior, and its presence is expected to bring a strong coupling between the J$_{eff}$ = 1/2 pseudospins and the lattice degrees of freedom\cite{Porras2019}. For instance, the J$_{eff}$ = 1/2 isospin directions are nearly locked to the rotation of IrO$_{6}$ octahedra along the \textit{c}-axis\cite{Samanta2018}.

In general, the coupling of magnetism to the ionic displacements in second order leads to a \textit{spin-phonon} coupling in magnetic materials, which may provide useful information on the microscopic exchange coupling mechanism\cite{Granado1999,Calder2015,Sohn2017}. In this letter, we investigate the temperature-dependent phonon Raman scattering of Sr$_{2}$IrO$_{4}$ single crystal. Our results reveal the existence of an \textit{isospin-phonon} coupling effect below the magnetic ordering temperature. We have also observed an asymmetric phonon lineshape of apical O stretching A$_{1g}$ mode, which is ascribed to the coupling of this phonon with a low-energy continuum of electronic excitations.   

The Sr$_{2}$IrO$_{4}$ single crystal was grown by the flux method. The base materials SrCO$_{3}$, IrO$_{2}$, and SrCl$_{2}$ (flux) were mixed in molar ratio 2:1:7, respectively; melted at 1280$^{\circ}$C in a platinum crucible, and then cooled down to 880$^{\circ}$C at the rate of 6$^{\circ}$C/hour. Finally, the crystals were free cooled to room temperature in a programmable furnace. The crystals were in sub-mm size (0.5-0.8 mm) with shiny surfaces. Raman scattering experiments were performed in a quasi-backscattering geometry using the 488.0 nm line of Ar$^{+}$ laser with focus spot size $\sim$ 50 $\mu$m. A Jobin-Yvon T64000 triple-mate spectrometer with 1800 mm$^{-1}$ gratings was employed. A LN$_{2}$-cooled multichannel CCD was used to collect and process the scattered data. The sample was mounted on a cold finger of closed-cycle He cryostat with base temperature 25K. The \textit{dc}-magnetization measurements as a function of temperature were performed with a Superconducting Quantum Interference Device (SQUID) magnetometer. The measurements were carried out under warming after field cooling (FCW, \textit{H} = 0.5 T).

\begin{figure}[ht]
	\centering
	\includegraphics[scale=0.65]{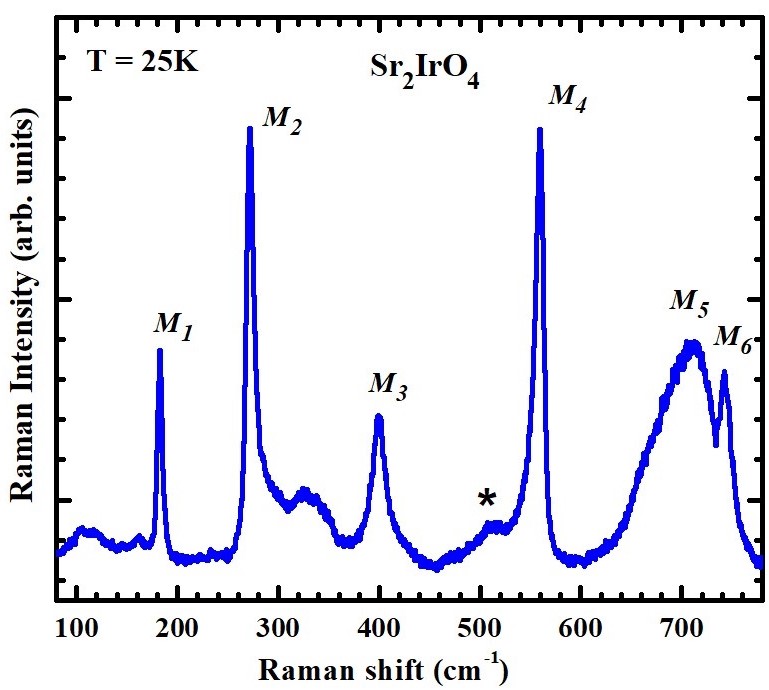}
	\begin{quotation}
		\caption[(Color online) Unpolarized Raman spectra of Sr$_{2}$IrO$_{4}$ single crystal at 25K. The phonon modes associated with M$_{1}$-M$_{6}$ are explained in the text.]{\small (Color online) Unpolarized Raman spectra of Sr$_{2}$IrO$_{4}$ single crystal at 25K. The phonon modes associated with M$_{1}$-M$_{6}$ are explained in the text.}
		\label{fig:Figure1}
	\end{quotation}
\end{figure}

Unpolarized Raman spectra of Sr$_{2}$IrO$_{4}$ single crystal at 25K are shown in Fig. \ref{fig:Figure1}. Six Raman-active optical modes have been detected and labeled as M$_{1}$ to M$_{6}$ at 183, 272, 399, 559, 715, and 742 cm$^{-1}$, respectively. These modes were identified in previous works\cite{Samanta2018,Gretarsson2017}. The most intense modes M$_{2}$ and M$_{4}$ at 25 K correspond to the A$_{1g}$ symmetry, where M$_{2}$ is a rotational mode of IrO$_{6}$ octahedra around the tetragonal \textit{c}-axis, and M$_{4}$ is a stretching of apical oxygen atoms along the \textit{c}-axis. 
The M$_{3}$ vibration with B$_{2g}$ symmetry is a bending mode of oxygen squares in the \textit{ab}-plane\cite{Samanta2018}. Peak M$_{1}$ is a superposition of A$_{1g}$ and B$_{2g}$ phonons with closely spaced frequencies. The A$_{1g}$ mode corresponds to a rotation of IrO$_{6}$ octahedra combined with an in-phase Sr displacement along \textit{c}-axis; while B$_{2g}$ is associated with the out-of-phase Sr vibrations\cite{Samanta2018}. The broad M$_{5}$ mode at $\sim$ 715 cm$^{-1}$ originates most likely from the two-phonon scattering\cite{Gretarsson2017}; while M$_{6}$ at $\sim$ 742 cm$^{-1}$ is a breathing mode of in-plane oxygen atoms with B$_{1g}$ symmetry\cite{Samanta2018,Cetin2012}. The observed modes are in good agreement with the previous symmetry-resolved single crystal Raman study\cite{Gretarsson2017,Cetin2012} and our \textit{ab}-initio lattice dynamics calculations and powder measurements\cite{Samanta2018}. The broad shoulder type peak ($\sim$ 334 cm$^{-1}$) at the high energy tail of M$_{2}$ was assigned as an A$_{1g}$ mode\cite{Gretarsson2017}, however we do not observe any change in frequency or intensity of this peak with temperature. The same type feature at $\sim$ 335 cm$^{-1}$ was also observed in Sr$_{2}$IrO$_{4}$ single crystal Raman scattering measurements\cite{Glamazda2014}. 
\begin{figure}[ht]
	\centering
	\includegraphics[scale=0.25]{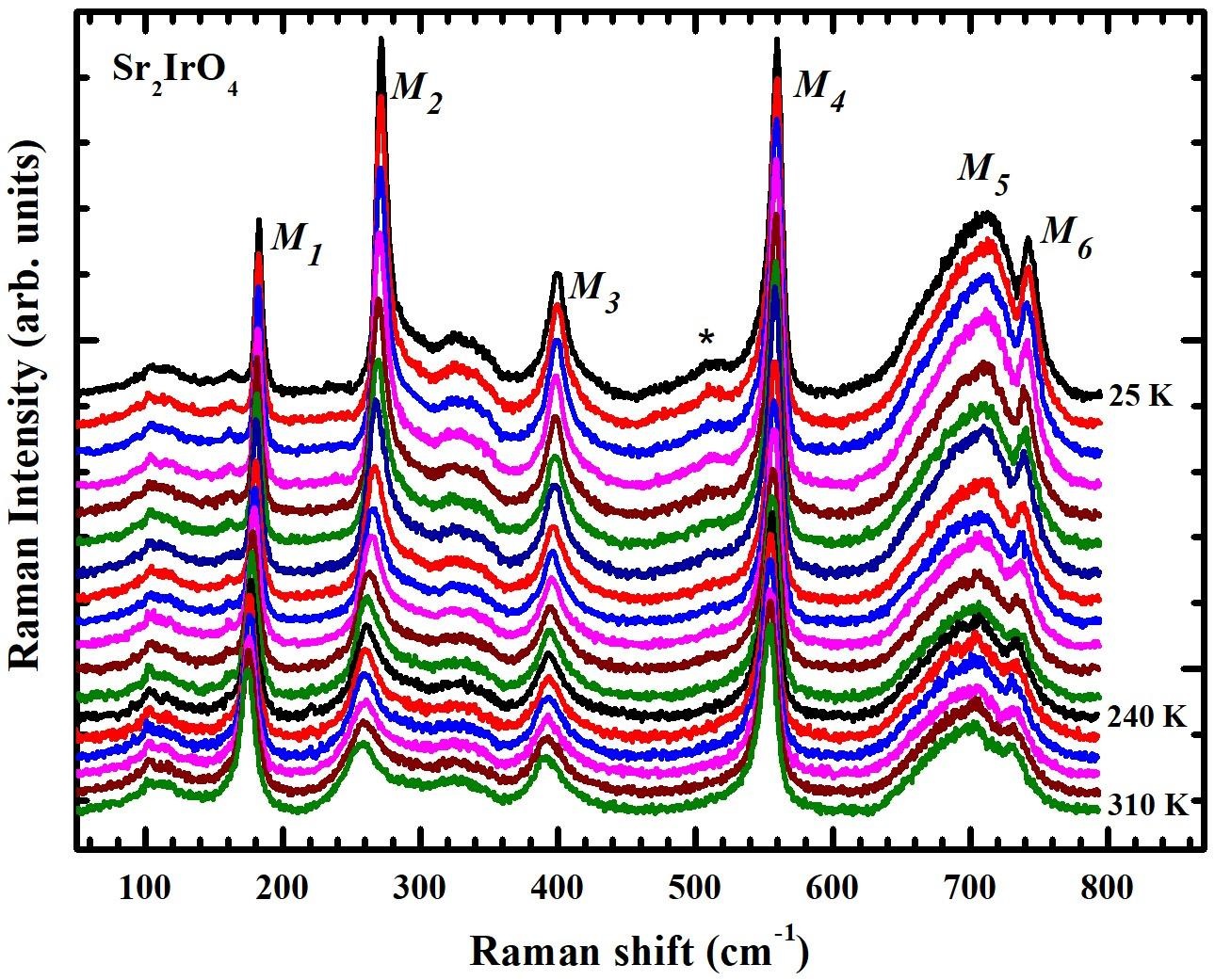}
	\begin{quotation}
		\caption[(Color online) Temperature dependent Raman spectra of  Sr$_{2}$IrO$_{4}$ single crystal. All the phonon modes are softening with increasing temperature.]{\small (Color online) Temperature dependent Raman spectra of  Sr$_{2}$IrO$_{4}$ single crystal. All the phonon modes are softening with increasing temperature.}
		\label{fig:Figure2}
	\end{quotation}
\end{figure}
Temperature dependent Raman scattering of Sr$_{2}$IrO$_{4}$ is shown in Fig. \ref{fig:Figure2}. All the phonon modes are shifted towards lower frequency with increasing temperature. M$_{4}$ (A$_{1g}$) shows an asymmetric lineshape throughout the temperature range under study, and the asymmetry decreases with increasing temperature. In addition to these phonon modes (M$_{1}$ to M$_{6}$), we have detected a shoulder-type peak (*) at $\sim$ 510 cm$^{-1}$, which gradually decreases its intensity and disappear at $\sim$ 240 K; interestingly, the frequency of this peak does not change with temperature. 

\begin{figure}[ht]
	\centering
	\includegraphics[scale=0.25]{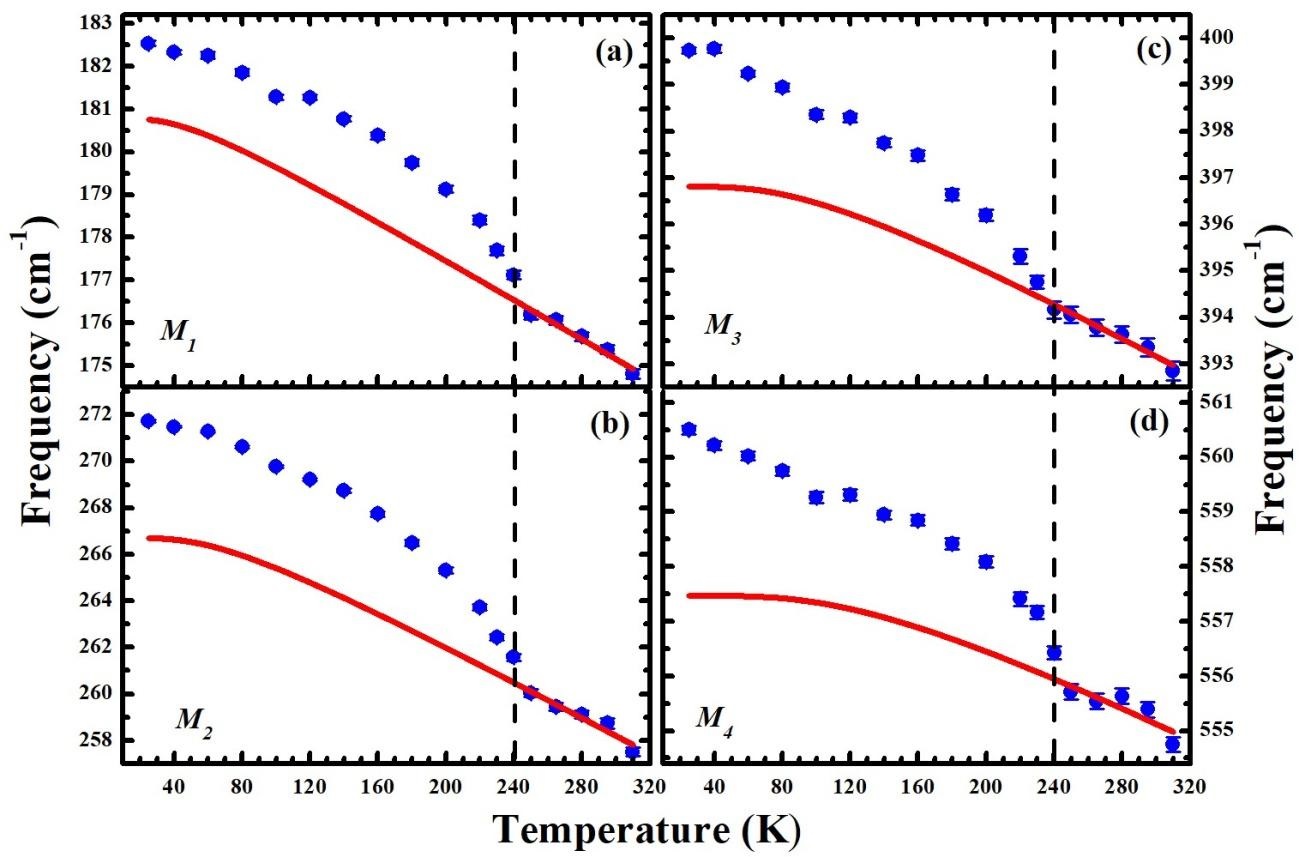}
	\begin{quotation}
		\caption[(Color online) (a)-(d) Solid circles represent the experimental phonon frequency change with temperature; solid lines represent the conventional anharmonic behavior estimated from two-phonon decay process.]{\small (Color online) (a)-(d) Solid circles represent the experimental phonon frequency change with temperature; solid lines represent the conventional anharmonic behavior estimated from two-phonon decay process.}
		\label{fig:Figure3}
	\end{quotation}
\end{figure}

\begin{figure}[ht]
	\centering
	\includegraphics[scale=0.25]{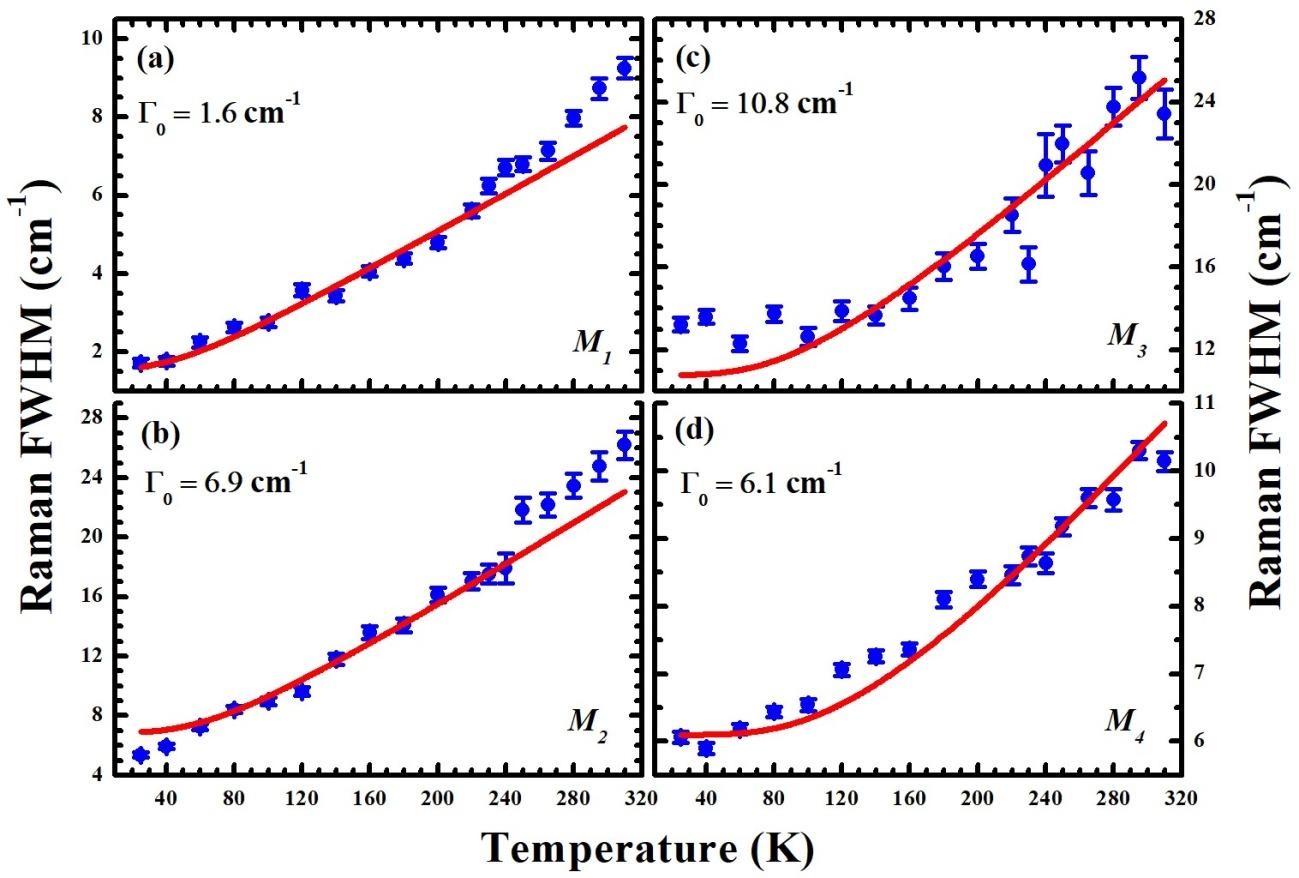}
	\begin{quotation}
		\caption[(Color online) (a)-(d) Solid circles represent the linewidth change as a function of temperature. The solid lines are the change in damping constant with temperature using two-phonon decay process.]{\small (Color online) (a)-(d) Solid circles represent the linewidth change as a function of temperature. The solid lines are the change in damping constant with temperature using two-phonon decay process.}
		\label{fig:Figure4}
	\end{quotation}
\end{figure}

\begin{figure}[ht]
	\centering
	\includegraphics[scale=0.33]{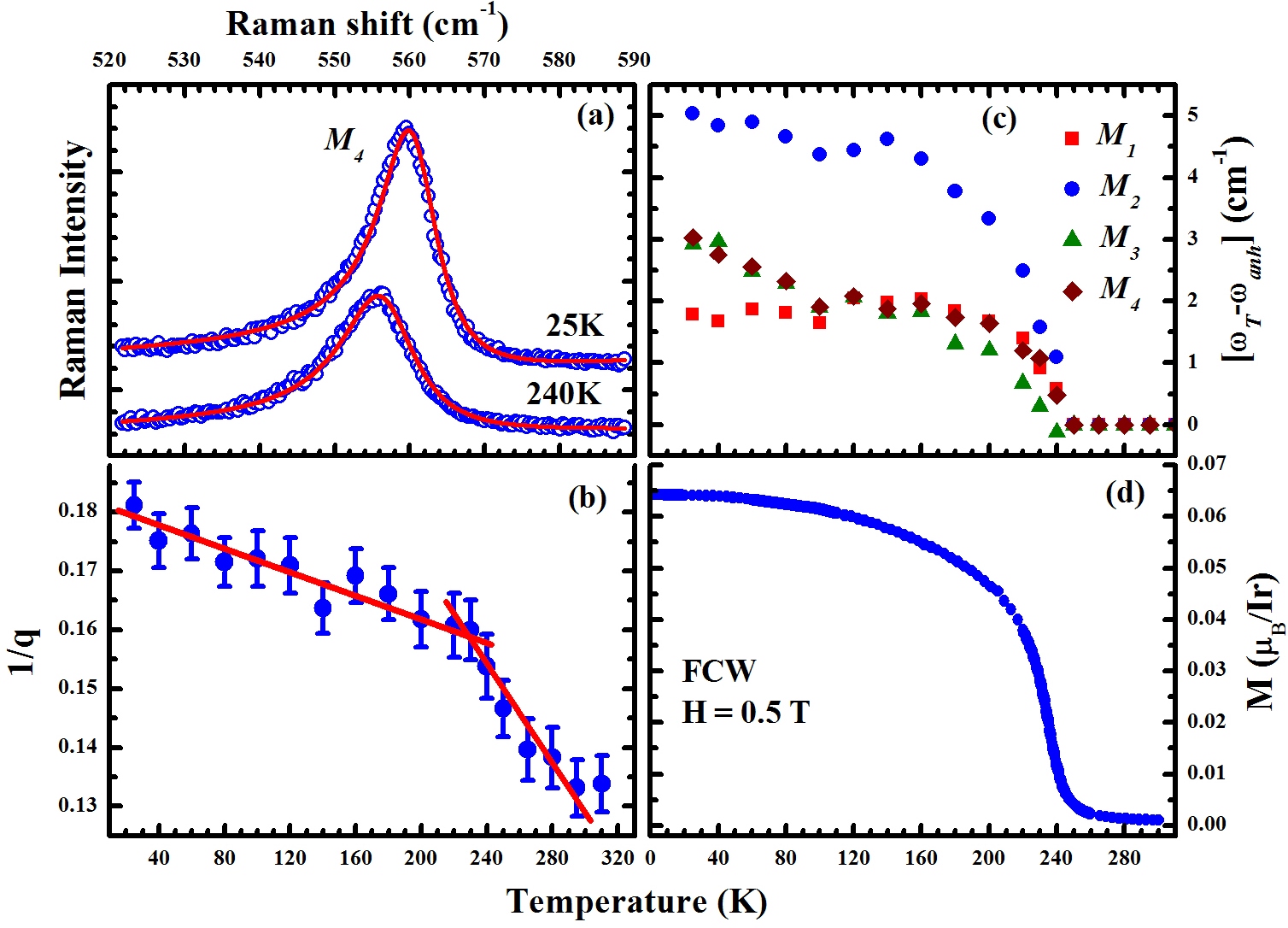}
		\begin{quotation}
			\caption[(Color online) (a) Asymmetric lineshape of M$_{4}$ (A$_{1g}$) mode, representing here for two temperature points 25 and 240K. Solid lines are the fitting with Fano-lineshape. (b) The phonon-continuum coupling strength as a function of temperature; the solid lines are to guiding the eye, the coupling strength shows an anomaly at $\sim$ 240 K. (c) Phonon frequencies ($\Delta\omega$) change due to the \textit{isospin-phonon} coupling as function of temperature. (d) Temperature dependent magnetization of bulk Sr$_{2}$IrO$_{4}$ in a field 0.5T.]{\small (Color online) (a) Asymmetric lineshape of M$_{4}$ (A$_{1g}$) mode, representing here for two temperature points 25 and 240 K. Solid lines are the fitting with Fano-lineshape. (b) The phonon-continuum coupling strength as a function of temperature; the solid lines are to guiding the eye, the coupling strength shows an anomaly at $\sim$ 240 K. (c) Phonon frequencies change ($\Delta\omega$) due to the \textit{isospin-phonon} coupling as function of temperature. (d) Temperature dependent magnetization of bulk Sr$_{2}$IrO$_{4}$ in a field 0.5T.}
			\label{fig:Figure5}
		\end{quotation}
\end{figure}

\begin{figure}[ht]
	\centering
	\includegraphics[scale=0.40]{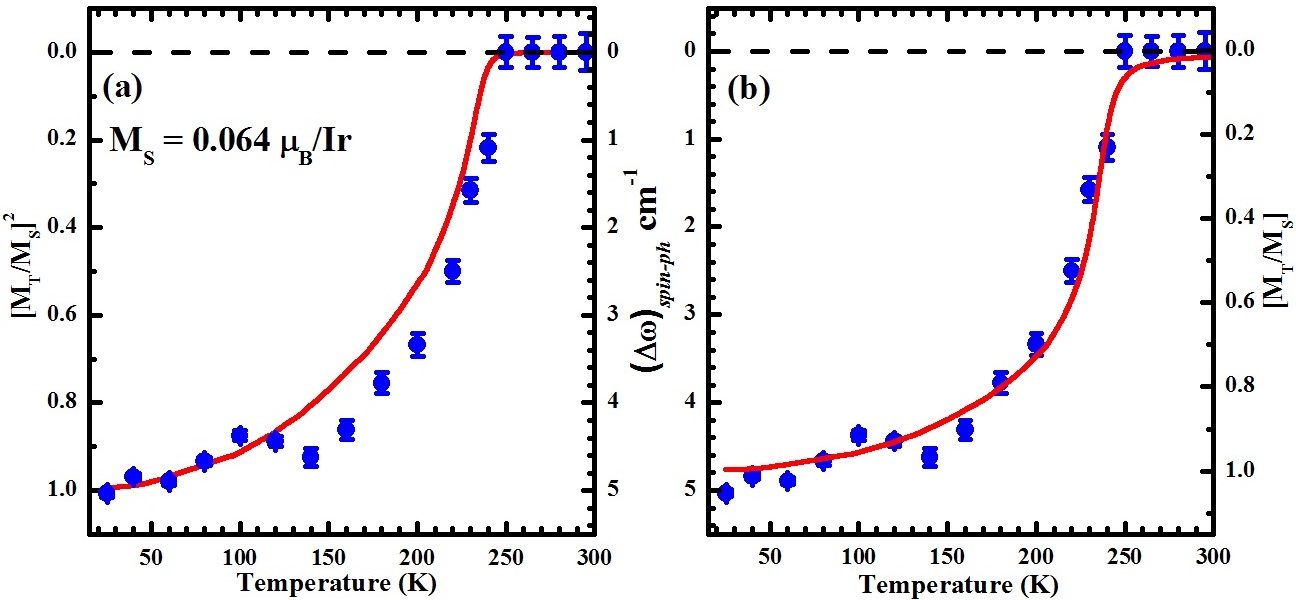}
	\begin{quotation}
		\caption[(Color online) (a) Represents the scaling of ($\Delta\omega$)$_{spin-ph}$ (solid circles) of M$_{2}$ mode with the \textit{normalized squire} of saturation magnetization (M$_{T}$/M$_{S}$)$^{2}$ (solid line); and (b) represents the scaling of ($\Delta\omega$)$_{spin-ph}$ with the \textit{linear normalized} saturation magnetization (M$_{T}$/M$_{S}$).]{\small (Color online) (a) Represents the scaling of ($\Delta\omega$)$_{spin-ph}$ (solid circles) of M$_{2}$ mode with the \textit{normalized squire} of saturation magnetization (M$_{T}$/M$_{S}$)$^{2}$ (solid line); and (b) represents the scaling of ($\Delta\omega$)$_{spin-ph}$ with the \textit{linear normalized} saturation magnetization (M$_{T}$/M$_{S}$).}
		\label{fig:Figure6}
	\end{quotation}
\end{figure}

Figures \ref{fig:Figure3} (a-d) show the change in M$_{1}$-M$_{4}$ phonon frequencies with temperature. For modes M$_{1}$-M$_{3}$, the phonon frequency and linewidth are obtained by fitting the phonon Raman profiles using a damped harmonic oscillator (DHO) function\cite{Menendez1984} $I(\omega) = [\chi_{0}\Gamma_{0}\omega\omega^{2}_{0}(n+1)]/[(\omega^{2}_{0} - \omega^{2})^{2} + \omega^{2}\Gamma^{2}_{0}]$, where $n = 1/[\exp(\hbar\omega/k_{B}T) - 1]$ is the phonon occupation number, $\omega_{0}$, $\Gamma_{0}$, and $\chi_{0}$ are the phonon frequency, linewidth, and intensity, respectively. The asymmetric phonon profile of M$_{4}$ mode is fitted with a Fano lineshape (see below). The solid lines represent the fitting of the experimental data using two phonon decay equation $\omega(T) = \omega_{0} - \Delta\omega_{anh}(T)$, where $\omega_{0}$ is the harmonic frequency of the mode when temperature approaches close to 0 K, and $\Delta\omega_{anh}(T)$ is the frequency shift due to the variation of self-energy induced by phonon-phonon interaction. Considering the two-phonon decay, the $\Delta\omega_{anh}(T)$ can be represented as $\Delta\omega_{anh}(T) = A\{1 + 2/[\exp(\hbar\omega/2k_{B}T) - 1]\}$\cite{Balkanski1983}. All the mode frequencies show an upturn below $\sim$ 240K, and deviate from the conventional anharmonic behavior of two-phonon decay process. We have also observed a fluctuation of phonon mode frequencies at $\sim$ 100 K.  
Figures \ref{fig:Figure4}(a-d) represent the temperature dependent damping constant $\Gamma_{anh}(T)$ of the modes M$_{1}$-M$_{4}$. The solid lines represent the expected $\Gamma_{anh}(T)$ curves of the modes according to the two-phonon decay $\Gamma_{anh}(T) = \Gamma_{0}\{1 + 2/[\exp(\hbar\omega/2k_{B}T) - 1]\}$\cite{Balkanski1983}. The inclusion of three or more phonon decay are expected to add only minor corrections in both frequency and damping constant with respect to the two-phonon decay in the temperature interval under study.  For M$_{1}$ and M$_{2}$, an increased broadening is observed above T$_{N}$ $\sim$ 240 K, indicating that isospin disorder contribute to the damping of these modes. 
Remarkably, M$_{4}$ shows an asymmetric lineshape in the whole temperature range (25-310K); Fig. \ref{fig:Figure5}(a) illustrates the phonon profile of M$_{4}$ at 25K and 240K at selected temperatures. The solid lines represent the fitting of asymmetric phonon profile using Fano line-shape $I(\omega) = I_{0}(q+\epsilon)/(1+\epsilon)^{2}$, where $I_{0}$ is the intensity, \textit{q} is the asymmetry parameter, and $\epsilon \equiv (\omega - \omega_{0})/\Gamma$, with $\omega_{0}$ and $\Gamma$ being the phonon frequency and linewidth, respectively\cite{Fano1961}. The inverse of asymmetry parameter (1/\textit{q}) is a measure of the electronic continuum-phonon coupling strength. It is interesting to note that the (1/\textit{q}) decreases slowly with increasing temperature up to T$_{N}$, and decrease steeply afterward (Fig. \ref{fig:Figure5}(b)). The anomalous phonon frequency change ($\Delta\omega$) of modes M$_{1}$-M$_{4}$, obtained after subtraction of the expected anharmonic contribution, is shown in Fig. \ref{fig:Figure5}(c).  The mode M$_{2}$ shows the maximum anomalous change ($\sim$ 5 cm$^{-1}$) at 25 K; whereas, all other modes this value is $\sim$ 2-3 cm$^{-1}$. The temperature-dependent bulk magnetization of Sr$_{2}$IrO$_{4}$ is shown in Fig. \ref{fig:Figure5}(d). 
It is well known that the phonon frequencies may be sensitive to spin correlations in magnetic materials. Anomalous phonon behavior below the magnetic ordering temperature is often interpreted as a manifestation of \textit{spin-phonon} coupling. This coupling occurs when the magnetic exchange energy is sensitive to a phonon normal coordinate, leading to the magnetic contribution to the harmonic energy of the lattice\cite{Garcia-Flores2012}. In J$_{eff}$ = 1/2 systems the magnetic Hamiltonian is written in terms of isospins rather than pure spins\cite{Jackeli2009}, and therefore the observed phonon anomalies may be ascribed to an \textit{isospin-phonon} coupling. Earlier, Gretarsson \textit{et al.} identified the \textit{pseudospin-lattice} coupling in Sr$_{2}$IrO$_{4}$ by analyzing the asymmetric lineshape of the lowest energy phonon mode ($\sim$ 187 cm$^{-1}$) above T$_{N}$, however, it was not possible to identify in their data anomalies in phonon frequency at T$_{N}$, that characterizes the spin-phonon coupling mechanism\cite{Gretarsson2016}. An experimental signature of the \textit{spin-phonon} coupling in antiferromagnetic materials in general is a scaling of the anomalous phonon frequency changes to the square of the sublattice magnetization below T$_{N}$\cite{Granado1999}, which in turn is proportional to the antiferromagnetic Bragg peak intensities obtained in magnetic diffraction measurements. In the present case, a complication arises from the non-collinear canted magnetic structure so that the dominant antiferromagnetic and the ferromagnetic sublattices may show different order parameters. In fact, a resonant x-ray scattering study shows that the temperature-dependence of the intensities of the antiferromagnetic Bragg peaks scales linearly rather than quadratically to the weak ferromagnetic moment obtained by bulk measurements\cite{Kim1329}, indicating a temperature-dependent spin canting angle. Thus, the anomalous phonon frequency changes due to the \textit{isospin-phonon} coupling in Sr$_{2}$IrO$_{4}$ may be also expected to scale linearly to the bulk magnetic moment. Indeed, Figs. \ref{fig:Figure6} (a) and \ref{fig:Figure6}(b) shows that ($\Delta\omega$)$_{spin-ph}$ scales better to the linear bulk magnetization M than to M$^{2}$. This analysis provide strong evidence that the anomalous phonon shifts reported here are due to an \textit{isospin-phonon} coupling mechanism. The additional anomaly of ($\Delta\omega$)$_{spin-ph}$ observed at $\sim$100K for the M$_{2}$ - M$_{4}$ modes (see Figs. \ref{fig:Figure3} and \ref{fig:Figure5}) may be due to a reorientation or change of the stacking pattern of the magnetic moments at this temperature\cite{Kim1329,Chikara2009,Ge2011,Franke2011}. 
A careful account of which normal modes of vibration show \textit{spin-phonon} couplings may provide valuable microscopic information on the magnetic coupling mechanisms. In our case, all studied modes M$_{1}$-M$_{4}$ show anomalous hardenings characteristic of \textit{spin-phonon} coupling. Modes M$_{1}$-M$_{3}$ are assigned to vibrations involving modulation of the Ir-O-Ir angle\cite{Samanta2018}, therefore \textit{isospin-phonon} anomalies are not too surprising to be observed for such modes. On the other hand, mode M$_{4}$ corresponds to a stretching of the apical oxygen against the Ir ions along the \textit{c}-direction\cite{Samanta2018}, and this mode might not be expected at first sight to modulate the magnetic coupling energy within the IrO$_{2}$ planes. We speculate that the \textit{isospin-phonon} coupling in this case takes place through an indirect mechanism involving a modulation of the tetragonal crystal-field splitting induced by this phonon, which in turn modulates the relative occupancy of the \textit{t$_{2g}$} orbitals and as a consequence the overall magnetic exchange energy.

As mentioned above the M$_{4}$ (A$_{1g}$) mode at $\sim$ 560 cm$^{-1}$ is a stretching mode of apical O along the \textit{c}-axis of IrO$_{6}$ octahedra; this phonon mode shows an asymmetric lineshape throughout the investigated temperature range. Such asymmetry is a signature of the coupling of the phonon with the underlying electronic charge and/or spin continuum excitation\cite{Gretarsson2017}. Earlier reports showed no indication of Fano-interference for the M$_{4}$ ($\sim$560 cm$^{-1}$) in either Sr$_{2}$IrO$_{4}$ or La-doped Sr$_{2}$IrO$_{4}$ single crystals\cite{Gretarsson2017}. On the other hand, an asymmetric phonon lineshape of M$_{1}$ ($\sim$185 cm$^{-1}$) mode was detected in Sr$_{2}$IrO$_{4}$ single crystal above T$_{N}$ ($\sim$240 K) and in La-doped Sr$_{2}$IrO$_{4}$ single crystal above 50 K\cite{Gretarsson2017}, which is not observed in our single crystal. In short, these results indicate low energy electronic excitations that are strongly sample dependent, and may be sensitive to the magnetic ordering temperature [see Fig. 5(b)]. Such effects are likely associated with the possible presence of charge carriers induced by intrinsic impurities in these samples. The macroscopic electrical properties of Sr$_{2}$IrO$_{4}$, including the non-ohmic \textit{I-V} response, is the subject of great interest in this field\cite{GCao2018}, and is most likely dominated by low energy electronic excitations such as those revealed by our spectroscopic data. Further theoretical and more systematic experimental studies are necessary to clarify this possible connection.  

In summary, temperature dependent Raman scattering of Sr$_{2}$IrO$_{4}$ single crystals was investigated, revealing new aspects of the interaction between the crystal lattice and electronic degrees of freedom. The anomalous phonon hardening observed below the magnetic transition temperature is a signature of \textit{isospin-phonon} coupling in this antiferromagnetic square lattice. The strongest effect is detected for the A$_{1g}$ ($\sim$ 272 cm$^{-1}$) mode associated with the modulation of in-plane Ir-O-Ir bond angle. The asymmetric lineshape of M$_{4}$ mode at 560 cm$^{-1}$ ($\sim$ 70 meV) originates from the coupling of this phonon with the underlying electronic continuum excitation, and this excitation is sensitive to the magnetic ordering temperature.

We thank Jean Souza for experimental support. This work was financially supported by FAPESP Grants No. 2016/00756-6 and No. 2017/10581-1, Brazil.

\newpage

\end{document}